\documentstyle[aps,epsfig,twocolumn]{revtex}

\begin{document}
\title{Cavity QED and quantum information processing with ``hot'' trapped
atoms}
\author{L.-M. Duan$^{1,2}$\thanks{
Email address: lmduan@caltech.edu}, A. Kuzmich$^{3}$, and H. J.
Kimble$^{3}$}
\address{$^{1}$Institute for Quantum Information, MC 107-81, California
Institute of Technology, Pasadena, CA 91125-8100\\
$^{2}$Laboratory of Quantum Information, USTC, Hefei 230026, China\\
$^{3}$Norman Bridge Laboratory of Physics 12-33, California
Institute of Technology, Pasadena, CA 91125} \maketitle

\begin{abstract}
We propose a method to implement cavity QED and quantum information
processing in high-Q cavities with a single trapped but non-localized atom.
The system is beyond the Lamb-Dick limit due to the atomic thermal motion.
Our method is based on adiabatic passages, which make the relevant dynamics
insensitive to the randomness of the atom position with an appropriate
interaction configuration. The validity of this method is demonstrated from
both approximate analytical calculations and exact numerical simulations. We
also discuss various applications of this method based on the current
experimental technology.

\end{abstract}

\section{Introduction}

Trapping of single atoms in high-Q cavities opens up exciting possibilities
for observation and manipulation of the dynamics of single \ particles and
for control of their interactions with single-mode photons \cite{1,2,3,4}.
Such possibilities could have wide applications, such as for generation of
non-classical or entangled optical pulses \cite{5,6}, for observing strong
cavity-QED effects \cite{3,4,7}, and more remarkably, for implementation of
quantum communication and computation \cite{8,8a,9,10,11}. The trapping
potential for confining single atoms can be created by diverse avenues,
including by the cavity QED light itself \cite{3,4}, by an additional
far-of-resonant trapping (FORT) beams \cite{2}, and by combining single
trapped ions with high-finesse optical cavities \cite{walther,blatt}. In
this paper, we will direct our attention principally to trapping in cavity
QED by way of an additional FORT beam, although our results are applicable
to broader settings.

The first experiment to achieve {\it strong coupling} in cavity QED with
trapped atoms was that of Ref.\cite{2}, which employed an intracavity FORT
beam and reported trapping lifetimes of $28$ms. By now, this experiment has
attained much longer trapping times, with recent work demonstrating
lifetimes in excess of $1$s\cite{icols01,mckeever02}. By contrast, atomic
localization by way of the cavity QED field itself has led to a trapping
within a single axial well with mean trapping time $\tau \approx 340\mu $s%
\cite{3} and to localization across many axial wells with mean time $\tau
\approx 280\mu $s\cite{4}.

The long trapping times achieved with an intracavity FORT beam set the stage
for diverse applications in Quantum Information Science, which motivates the
current analysis. However, one of the main obstacles to the experimental
demonstration of these applications is that the position of the trapped atom
is not well fixed within the cavity. The coupling rate $g$ between the
atomic internal levels and the cavity mode depends on the atom position $%
{\bf r}$ through the relation%
\begin{equation}
g\left( {\bf r}\right) =g_{0}\chi \left( {\bf r}\right)   \label{1e}
\end{equation}%
with the mode function%
\begin{equation}
\chi \left( {\bf r}\right) =\sin \left( k_{0}z\right) \exp \left[ -\left(
x^{2}+y^{2}\right) /w_{0}^{2}\right] ,  \label{2e}
\end{equation}%
where $g_{0}$ is the peak coupling rate, $w_{0}$ and $k_{0}=2\pi /\lambda
_{0}$ are respectively the width and the wave vector of the Gaussian cavity
mode, and $z$ is assumed to be along the axis of the cavity. Due to the
randomness of the atom position ${\bf r}$, we have an unknown randomly
changing coupling rate $g\left( {\bf r}\right) $. Most of the applications
of this setup assumed a fixed known coupling rate $g$. Therefore, before the
experimental demonstration of these schemes, first one needs to solve the
problem associated with the random coupling.

Intense experimental efforts have been taken to localize the atom inside the
cavity so as to fix the coupling rate $g\left( {\bf r}\right) $, with
notable recent success attained via ion traps \cite{walther,blatt}. In the
cavity QED experiments employing cold atoms and without FORT beams \cite%
{1,16,17}, atoms were dropped through the cavity and followed random
trajectories with large axial heating. As a result, the magnitude and the
sign of $g\left( {\bf r}\right) $ were not well controlled. With a FORT beam
and with current experimental capabilities\cite{2,icols01,mckeever02}, an
atom can be trapped inside one potential well along the cavity axis with a
fixed sign of $g\left( {\bf r}\right) $. But the atom still has appreciable
kinetic energy and is not fully localized, leading to significant variations
in the magnitude of the coupling rate $g\left( {\bf r}\right) $.

The randomness of the coupling rate $g\left( {\bf r}\right) $ comes from
several contributions: first, the trapped atom is still quite hot in the
current experimental setup. Its kinetic energy from the thermal motion is
typically lower but not much lower than the depth of the trapping potential.
The atom's oscillation amplitude $d$ in the trap is comparable to the
optical wave length $\lambda _{0}$, so it does not satisfy the usually
assumed Lamb-Dick condition $d\ll \lambda _{0}$. Due to the thermal motion
of the atom, the coupling rate $g\left( {\bf r}\right) $ typically has a
variation within a fact of $2$ with the current experimental technique.
Certainly, the atom will become better localized as cooling techniques are
adapted to cavity QED and its energy is reduced\cite{12,13}. However, due to
the presence of the cavity and the trapping potential, it is still
experimentally hard to achieve efficient cooling inside the cavity \cite%
{12,13}. Furthermore, even if we assume that the atom has been pre-cooled
and localized initially to the Lamb-Dick limit, the implemented application
protocols will still tend to heat the atom due to photon recoils from the
spontaneous emissions \cite{13',13''}. As a result of the heating, the atom
may go out of the Lamb-Dick limit after a short time. Finally, even if we
neglect all the motional and the heating effects of the trapped atom, there
is still some uncertainty of the coupling rate. The intracavity field of the
FORT beam forms many potential wells inside the cavity, and in current
experiments, one can not control and does not know precisely in which well
the atom is trapped. The FORT beam has a wavelength $\lambda _{F}$ different
from the cavity QED wavelength $\lambda _{0}$, so, even if the atom is kept
to be very cold and well localized at the bottom of the trapping potential
well, we still might not know exactly the coupling rate since the bottoms of
different potential wells have different coupling rates \cite{vanenk}.

Here, to overcome these difficulties, we propose a method to do cavity QED
and quantum information processing directly with hot atoms with an
inhomogeneous distribution in position and/or time varying locations. The
method is based on adiabatic passages with an appropriate interaction
configuration. Normally, schemes based on adiabatic passages are more
insensitive to certain parameter changes compared with the corresponding
Raman schemes. Some initial indication of insensitivity of the adiabatic
passage scheme to certain parameter changes was already illustrated in \cite%
{15'} for some cavity QED scheme. However, to make the whole system dynamics
insensitive to variations of the coupling rate $g\left( {\bf r}\right) $, we
also need to design an appropriate interaction configuration. The relevant
dynamics of adiabatic passages are determined by the relative ratio between
different coupling rates, and are almost independent of their absolute
values. Thanks to this property, with an appropriate design of the
interaction configuration, we can make different coupling rates have the
same dependence on the atom position ${\bf r}$, and therefore the system
dynamics determined by their relative ratios will become independent of $%
{\bf r}$. As a result, though the atom position may be unknown and time
dependent, the output signal from the cavity is still well controllable and
has definitely known properties. Note that the method described here is
different to some previous quantum computation schemes with hot trapped ions %
\cite{14,15}, where the Lamb-Dick condition is still required.

The paper is arranged as follows: In Sec. II, we explain the basic idea of
the method, and then describe and solve the model Hamiltonian analytically
within the adiabatic approximation. In Sec. III, we give an exact numerical
simulation of the model, with the emphasis on checking the validity of the
introduced approximations and calculating various kinds of noise magnitudes
relevant for the on-going experimental efforts. The calculations show that
we can get reasonably good signal-to-noise ratios with typical experimental
values for the parameters. In Sec. IV, we discuss various applications of
this method. By incorporating this method into some previously-known
schemes, we show that with hot non-localized atoms, one can still realize
many kinds of cavity QED and quantum information processing schemes,
including, for instance, the controllable single-photon or entangled photon
source, quantum communication between cavities, atomic entanglement
generation, teleportation, and Bell inequality detection. Section 5 gives a
synopsis of parameters relevant to our current experiment for a single atom
trapping with a FORT beam at Caltech \cite{2,icols01,mckeever02}. We
summarize the results in the final section.

\section{Cavity QED with a non-localized trapped atom: the scheme}

\subsection{Basic idea}

First, we explain the basic idea of this method by considering a single
trapped atom, which has three effective levels $\left| g\right\rangle $, $%
\left| e\right\rangle $, $\left| s\right\rangle $, as shown in Fig. 1. The
two ground states $\left| g\right\rangle $ and $\left| s\right\rangle $ can
correspond, for instance, to sub-Zeeman levels in the $F=3$ and $F=4$
manifold respectively for the cesium atom. The transition $\left|
e\right\rangle \rightarrow \left| s\right\rangle $ is coupled resonantly to
the cavity QED mode $a$ with a coupling rate $g\left( {\bf r}\right) $ in
the form of Eq. (1). A classical laser field $\varepsilon \left( t\right) $
incident from one mirror of the cavity (see Fig. 1) drives the transition $%
\left| g\right\rangle \rightarrow \left| e\right\rangle $ through another
cavity mode $a^{\prime }$. We assume for simplicity that $a$ and $a^{\prime }
$ have the same spatial mode structure with the same frequency (for example,
they can be of different polarizations). The driving laser $\varepsilon
\left( t\right) $ is resonant to the transition $\left| g\right\rangle
\rightarrow \left| e\right\rangle $, so it is far-of-resonant to the cavity
mode $a^{\prime }$ with a large detuning $\omega _{gs}$, where $\omega _{gs}$
denotes the splitting between the levels $\left| g\right\rangle $ and $%
\left| s\right\rangle $. Due to the off-resonant driving by $\varepsilon
\left( t\right) $, $a^{\prime }$ can be described classically by its mean
value $\left\langle a^{\prime }\right\rangle =\alpha \left( t\right)
e^{-i\omega _{ge}t}$ ($\omega _{ge}$ is the frequency splitting between the
levels $\left| g\right\rangle $ and $\left| e\right\rangle $), which couples
resonantly to the transition $\left| g\right\rangle \rightarrow \left|
e\right\rangle $ with a Rabi oscillation frequency $\Omega \left( {\bf r}%
,t\right) $. Since $a$ and $a^{\prime }$ have the same spatial mode
structure, the Rabi frequency $\Omega \left( {\bf r},t\right) $ will depend
on the the atom position ${\bf r}$ by the same mode function $\chi \left(
{\bf r}\right) $, i.e., $\Omega \left( {\bf r},t\right) $ can be factorized
as%
\begin{equation}
\Omega \left( {\bf r},t\right) =\Omega _{0}\left( t\right) \chi \left( {\bf r%
}\right) =r_{o}g_{0}\alpha \left( t\right) \chi \left( {\bf r}\right)
\label{2}
\end{equation}%
where $r_{o}$ represents the fixed ratio of the\ Clebsch-Gordan coefficients
for the transitions $\left| g\right\rangle \rightarrow \left| e\right\rangle
$ and $\left| s\right\rangle \rightarrow \left| e\right\rangle $.

\begin{figure}[tbp]
\epsfig{file=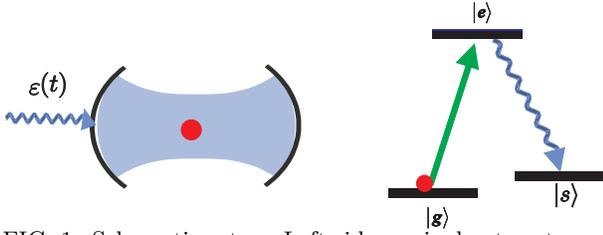,width=8cm} \caption{Schematic setup. Left
side: a single atom trapped in a high-Q
cavity, which is driven by a classical laser pulse $\protect\varepsilon %
\left( t\right) $. Right side: the relevant atomic level structure.}
\end{figure}

To understand the basic idea of this method, let us first look at a very
simplified picture by neglecting the coupling of the mode $a$ to the cavity
output. The system is then described by the following simple Hamiltonian in
the rotating frame (setting $\hbar =1$)

\begin{equation}
H_{\text{sim}}=\Omega \left( {\bf r},t\right) \sigma _{eg}+g\left( {\bf r}%
\right) a\sigma _{es}+H.c.  \label{3}
\end{equation}%
where $\sigma _{\mu \nu }=\left| \mu \right\rangle \left\langle \nu \right| $
$\left( \mu ,\nu =g,e,s\right) $\ are the atomic transition operators, and $%
H.c.$ stands for the Hermitian conjugate. The Hamiltonian $H_{\text{sim}}$
has the well-known dark state $\left| D\right\rangle $ (the instantaneous
eigenstate with a zero eigenvalue) with the form \cite{15'}%
\begin{eqnarray}
\left| D\right\rangle &=&\frac{1}{\sqrt{\left| g\left( {\bf r}\right)
\right| ^{2}+\left| \Omega \left( {\bf r},t\right) \right| ^{2}}}\left(
g\left( {\bf r}\right) \left| g\right\rangle \left| 0\right\rangle -\Omega
\left( {\bf r},t\right) \left| s\right\rangle \left| 1\right\rangle \right)
\nonumber \\
&=&\frac{1}{\sqrt{1+\left| r_{o}\alpha \left( t\right) \right| ^{2}}}\left(
\left| g\right\rangle \left| 0\right\rangle -r_{o}\alpha \left( t\right)
\left| s\right\rangle \left| 1\right\rangle \right) ,  \label{4}
\end{eqnarray}%
where $\left| 0\right\rangle $ and $\left| 1\right\rangle $ represent the
zero and the one photon state of the cavity mode $a$. Note that the dark
state $\left| D\right\rangle $ actually only depends on the ratio between
the parameters $g\left( {\bf r}\right) $ and $\Omega \left( {\bf r},t\right)
$, so it becomes independent of the random atom position ${\bf r}$ with the
interaction configuration specified above. If we start with the atom in the
ground state $\left| g\right\rangle $, and gradually increase the Rabi
frequency $\Omega \left( {\bf r},t\right) $, under the adiabatic
approximation, the state of the system will remain in the dark state $\left|
D\right\rangle $, which gradually evolves into the final state $\left|
s\right\rangle \left| 1\right\rangle $. Due to the independence of the state
$\left| D\right\rangle $ on the variable ${\bf r}$, the relevant dynamics of
this adiabatic evolution also becomes independent of the random atom site $%
{\bf r}$. This is the basic idea of the method to eliminate the influence of
the randomness in the coupling coefficient $g\left( {\bf r}\right) $.

Note that to make the dark state and the relevant dynamics independent of
the random atom position ${\bf r}$, the driving pulse and the cavity mode
need to have the same spatial mode structure. This is why the classical
driving pulse is matched to spatial mode of the cavity field, both along the
cavity axis and transversely, which is routinely accomplished by way of
illumination from one side mirror of the cavity. This configuration is
different from the original proposals for adiabatic dynamics in cavity QED %
\cite{15'} in which the propagation direction of the driving pulse is
perpendicular to the cavity axis with uniform illumination intensity. It is
also distinct from the configuration employed in some recent interesting
experiments directed toward achieving a single-photon source \cite{16,17},
which likewise employed uniform illumination transverse to the cavity axis
and for which the atom is not localized axially. As a result, in these
experiments some of the dynamics, such as the output pulse shape and phase,
still depend on the unknown position of the randomly moving atom, and are
thus not fully controllable and reversible, as has been seen from the
experiments.

To guarantee an adiabatic evolution, we need to fulfill the adiabatic
condition, which means the evolution time $T$ should be significantly longer
than the frequency gap $\delta $ between the dark state and some other
eigenstates of the Hamiltonian $H_{\text{sim}}$. The error probability due
to the non-adiabaticity is estimated by $p_{ad}=1/\left( \delta T\right)
^{2} $. For the Hamiltonian $H_{\text{sim}}$, the frequency gap $\delta $ is
given by $\delta =\sqrt{\left| g\left( {\bf r}\right) \right| ^{2}+\left|
\Omega \left( {\bf r},t\right) \right| ^{2}}$. Thus, the adiabatic condition
$\left[ \left| g\left( {\bf r}\right) \right| ^{2}+\left| \Omega \left( {\bf %
r},t\right) \right| ^{2}\right] T^{2}\gg 1$ depends on the atom position $%
{\bf r}$. If the coupling coefficient $g\left( {\bf r}\right) $ changes by a
factor of $2$, the error probability $p_{ad}$ will change by a factor of $4$
for the same evolution time $T$. However, if $T$ is sufficiently long, the
error probability $p_{ad}$ remains to be small, and the relevant system
dynamics will be still very insensitive to the randomness of the atom
position. To estimate $p_{ad}$, we can use the average value of the coupling
rate $g\left( {\bf r}\right) $.

In the above simple picture, we neglect the coupling of the mode $a$ to the
cavity output. This is only a valid picture in the good-cavity limit with
the evolution time $T\ll 1/\kappa $, where $\kappa $ is the cavity decay
rate. However, in practice, it is better to operate the system in the limit
with $T\geq 1/\kappa $. There are several advantages of operating the system
in this limit: first, without the requirement $T\ll 1/\kappa $, \ it is
easier to satisfy the adiabatic condition for which $T$ should be
sufficiently long; second, in this limit it is easier to modulate the Rabi
frequency $\Omega \left( {\bf r},t\right) $ by changing the intensity of the
driving laser $\varepsilon \left( t\right) $ incident from one side mirror
of the cavity. In this way, one can efficiently control the pulse shape of
the cavity output by modulating the shape $\varepsilon \left( t\right) $ of
the driving laser, which is useful for many applications. In the limit $%
T\geq 1/\kappa $, we need to take into account from the beginning the
coupling of the mode $a$ to the continuum cavity output, and the whole
system will then have infinite levels. We will describe in the next
subsection this more involved interaction configuration. The above simple
three-level picture, though it does not describe the real experimental
configuration, does help for the understanding of the basic idea of the
adiabatic method.

\subsection{Theoretical model and its approximate analytical solution}

Now we look at the more complicate theoretical model, which includes the
coupling of the mode $a$ to the continuum cavity output. If we adiabatically
apply a classical diving pulse $\varepsilon \left( t\right) $ as shown in
Fig. 1, one photon will be emitted from the transition $\left|
e\right\rangle \rightarrow \left| s\right\rangle $, and the cavity will
output a single-photon pulse. We want to show in the below that this
single-photon pulse has a definite pulse shape which is independent of the
randomness in the atom position ${\bf r}$ and in the coupling rate $g\left(
{\bf r}\right) $. In this way, although the atom position and the absolute
value of the light-atom coupling rate are not fully controlled, we can
nevertheless fully control the properties of the output single-photon pulse
by modulating the driving laser pulse $\varepsilon \left( t\right) $. This
is an important feature for many applications of this setup, which we will
discuss in Sec. IV. There are several equivalent ways to describe the
coupling of the mode $a$ to the continuum cavity output. Since we want to
calculate the output pulse shape within the adiabatic approximation, it is
convenient to use the Hamiltonian approach \cite{18,19}. The whole
Hamiltonian, including the coupling to the cavity output, has the following
form in the rotating frame \cite{19}%
\begin{eqnarray}
H &=&(\Delta -i\gamma _{s}/2)\sigma _{ee}+\left[ \Omega \left( {\bf r}%
,t\right) \sigma _{eg}+g\left( {\bf r}\right) a\sigma _{es}+H.c.\right]
\nonumber \\
&&+i\sqrt{\kappa /2\pi }\int_{-\omega _{b}}^{+\omega _{b}}d\omega \left[
a^{\dagger }b\left( \omega \right) -ab^{\dagger }\left( \omega \right) %
\right]  \label{5} \\
&&+\int_{-\omega _{b}}^{+\omega _{b}}d\omega \left[ \omega b^{\dagger
}\left( \omega \right) b\left( \omega \right) \right] ,  \nonumber
\end{eqnarray}%
where $b\left( \omega \right) $, with the standard commutation relation $%
\left[ b\left( \omega \right) ,b^{\dagger }\left( \omega ^{\prime }\right) %
\right] =\delta \left( \omega -\omega ^{\prime }\right) $, denote the
one-dimensional free-space modes which couple to the cavity mode $a$. We
only need to consider the free-space modes within a finite bandwidth $\left[
\omega _{se}-\omega _{b},\omega _{se}+\omega _{b}\right] $ with the carrier
frequency $\omega _{se}$ ($\omega _{se}$ is the frequency splitting between
the levels $\left| s\right\rangle $ and $\left| e\right\rangle $), since all
the modes outside of this bandwidth have negligible contributions to the
dynamics due to the large detuning (larger than $\omega _{b}$). Within this
bandwidth, the coupling between $b\left( \omega \right) $\ and the cavity
mode $a$ is approximately a constant, and we denote it by $\sqrt{\kappa
/2\pi }$ for convenience, where $\kappa $ is the effective cavity decay
rate, as we will see. The bandwidth $\omega _{b}$ should be chosen to be
much larger than $\kappa $, but still much smaller than $\omega _{se}$.

We have assumed that the driving laser and the cavity mode $a$ couple
resonantly to the corresponding free-space atomic transitions. However, we
emphasize that our scheme still works for the case of off-resonant coupling.
By considering the off-resonant scheme, there is no win with respect to
losses due to the atomic decay, since in this case the time scale also slows
down. So it suffices here to consider the resonant coupling case. However,
in the Hamiltonian (6), it is still helpful to include a
single-photon-transition detuning $\Delta $ to account for the trapping
potential difference for the levels $\left| g\right\rangle $ and $\left|
e\right\rangle $ induced by the FORT beam (this potential is basically the
same for the levels $\left| s\right\rangle $ and $\left| g\right\rangle $
for a FORT beam with linear polarization as in our current experiments). The
potential difference between the level $\left| g\right\rangle $ and $\left|
e\right\rangle $ in general depends as well on the random atom position $%
{\bf r}$.

The imaginary part of the Hamiltonian (6) accounts for the spontaneous
emission loss, where $\gamma _{s}$ denotes the total spontaneous emission
rate of the upper level $\left| e\right\rangle $. In writing this form, we
have assumed that the spontaneous emission photon escapes and that the atom
after a spontaneous emission will not be repumped. This is a good assumption
for the interesting region where the spontaneous emission loss is not big,
and the atom thus has a very small probability to be repumped after emitting
a spontaneous emission photon. As a result of this assumption, the
spontaneous emission only contributes to the leakage error which is properly
represented by Eq. (6) \cite{19ab}.

We treat the atom position ${\bf r}$ in the Hamiltonian (6) as a classical
stochastic variable, and neglect its quantum nature. This is a good
approximation for the current experimental situation where the atom is still
quite hot. There have been some analysis of the noise from quantum motional
effects in high-Q cavities with very cold atoms \cite{19a}.

We start with the atom in the ground state $\left| g\right\rangle $, and
then apply a classical driving pulse $\varepsilon \left( t\right) $. This
pulse can efficiently control the time evolution of the Rabi frequency $%
\Omega \left( {\bf r},t\right) $ in the Hamiltonian (6). To see this, we
write the input-output equation for the cavity mode $a\prime $ \cite{19}%
\begin{equation}
\stackrel{.}{a\prime }=-i\omega _{se}a\prime -\frac{\kappa }{2}a\prime -%
\sqrt{\kappa }a_{\text{in}}^{\prime }\left( t\right) ,  \label{6}
\end{equation}%
where $a_{\text{in}}^{\prime }\left( t\right) $ is the field operator for
the input driving pulse coupling to the mode $a\prime $, with $\left\langle
a_{\text{in}}^{\prime }\left( t\right) \right\rangle =\varepsilon \left(
t\right) $\ and $\left[ a_{\text{in}}^{\prime }\left( t\right) ,a_{\text{in}%
}^{\prime \dagger }\left( t^{\prime }\right) \right] =\delta \left(
t-t^{\prime }\right) $. By assumption, the mode $a\prime $ has the same
frequency as the mode $a$ which is resonant to the free-space atomic
transition $\left| s\right\rangle \rightarrow \left| e\right\rangle $, so
the eigen-frequency of $a\prime $ is $\omega _{se}$. Such a situation
corresponds, for example, to the case of the $(a,a\prime )$ modes of
orthogonal polarization, but degenerate in frequency, although this is not
an essential requirement. In Eq. (7), we have neglected the small depletion
of $a\prime $\ caused by the coupling to the atomic transition $\sigma _{eg}$
since $a\prime $ is driven by a strong classical pulse $\varepsilon \left(
t\right) $ which dominates its time evolution. We write the mean values of $%
a^{\prime }$ and $a_{in}^{\prime }\left( t\right) $ as $\left\langle
a^{\prime }\right\rangle =\alpha \left( t\right) e^{-i\omega _{ge}t}$ and $%
\left\langle a_{\text{in}}^{\prime }\left( t\right) \right\rangle
=\varepsilon \left( t\right) =\widetilde{\varepsilon }\left( t\right)
e^{-i\omega _{ge}t}$, where $\widetilde{\varepsilon }\left( t\right) $ is
the slowly-varying amplitude of the driving laser. Form Eq. (7), we get a
time evolution equation for the mean value $\alpha \left( t\right) $, which
has the following immediate solution%
\begin{equation}
\alpha \left( t\right) =\int_{0}^{t}\widetilde{\varepsilon }\left( \tau
\right) e^{(i\omega _{gs}-\kappa /2)\left( t-\tau \right) }d\tau .  \label{7}
\end{equation}%
The variation rate of $\widetilde{\varepsilon }\left( \tau \right) $ is
characterized by the inverse of the operation time $T$ (the pulse duration),
which is typically much smaller than the hyperfine frequency splitting $%
\omega _{gs}$ (about $9$GHz for cesium atoms). Hence, a partial integration
of Eq. (8) yields%
\begin{equation}
\alpha \left( t\right) \simeq \frac{\widetilde{\varepsilon }\left( t\right)
-e^{(i\omega _{gs}-\kappa /2)t}\widetilde{\varepsilon }\left( 0\right) }{%
-i\omega _{gs}+\kappa /2}\left[ 1+o\left( \frac{1}{\omega _{gs}T}\right) %
\right] .  \label{8}
\end{equation}%
We assume that $\widetilde{\varepsilon }\left( t\right) $ gradually
increases from zero with $\widetilde{\varepsilon }\left( 0\right) \simeq 0$.
Then, within a good approximation, we have $\alpha \left( t\right) \propto
\widetilde{\varepsilon }\left( t\right) $ from Eq. (9). In the following,
without loss of generality, we assume $\alpha \left( t\right) $ to be real
by choosing an appropriate constant phase of $\widetilde{\varepsilon }\left(
t\right) $. The time behavior of the Rabi frequency $\Omega \left( {\bf r}%
,t\right) $ is completely determined by $\alpha \left( t\right) $ (note that
$\Omega \left( {\bf r},t\right) =r_{o}g_{0}\alpha \left( t\right) \chi
\left( {\bf r}\right) $ from Eq. (3)), that is, by the amplitude $\widetilde{%
\varepsilon }\left( t\right) $ of the driving laser.

The dark state (5) can be rewritten as $\left| D\right\rangle =\cos \theta
\left| g\right\rangle \left| 0\right\rangle -\sin \theta \left|
s\right\rangle \left| 1\right\rangle $, with $\cos \theta =1/\sqrt{1+\left|
r_{o}\alpha \left( t\right) \right| ^{2}}$ independent of the atom position $%
{\bf r}$. The state $\left| B\right\rangle $ complementary to the dark state
is usually called the bright state with $\left| B\right\rangle =\sin \theta
\left| g\right\rangle \left| 0\right\rangle +\cos \theta \left|
s\right\rangle \left| 1\right\rangle $. To solve the dynamics governed by
the Hamiltonian (6), we can expand the state $\left| \Psi \right\rangle $ of
the whole system into the following superposition%
\begin{equation}
\left| \Psi \right\rangle =\left( c_{d}\left| D\right\rangle +c_{b}\left|
B\right\rangle +c_{e}\left| e\right\rangle \left| 0\right\rangle \right)
\otimes \left| \text{vac}\right\rangle +\left| s\right\rangle \left|
0\right\rangle \otimes \left| \varphi _{1}\right\rangle ,  \label{9}
\end{equation}%
where $\left| \text{vac}\right\rangle $ denotes the vacuum state of the
free-space modes $b\left( \omega \right) $, and
\begin{equation}
\left| \varphi _{1}\right\rangle =\int_{-\omega _{b}}^{+\omega _{b}}d\omega
c_{\omega }b^{\dagger }\left( \omega \right) \left| \text{vac}\right\rangle
\label{phi1}
\end{equation}%
represents the state (not-normalized) of the single-photon output pulse. The
coefficients $c_{d},$ $c_{b},$ $c_{e}$\ and $c_{\omega }$ in Eq. (10) are
time dependent. At the time $t=0$, we have $c_{d}=1$, $c_{b}=c_{e}=c_{\omega
}=0$ and $\cos \theta =1$. After applying a classical driving pulse $%
\varepsilon \left( t\right) $, $\cos \theta $ slowly changes with $\alpha
\left( t\right) $, and we need to compute the time evolution of all the
coefficients $c_{d},c_{b},c_{e},c_{\omega }$ in Eq. (10) by substituting $%
\left| \Psi \right\rangle $ into the Schroedinger equation $i\partial
_{t}\left| \Psi \right\rangle =H\left| \Psi \right\rangle $.

To go on with this task, let us first take the adiabatic approximation,
which assumes the time derivative $\partial _{t}\cos \theta \approx 0$. As a
result, $\partial _{t}\left| D\right\rangle $ and $\partial _{t}\left|
B\right\rangle $ become negligible. We will check the validity of the
adiabatic approximation and calculate various non-adiabatic corrections in
the next section through numerical methods. In the adiabatic limit, the
populations in the bright state $\left| B\right\rangle $ and in the excited
state $\left| e\right\rangle $ are negligible, so we assume $c_{b}\approx
c_{e}\approx 0$. The coefficients $c_{d}$\ and $c_{\omega }$ satisfy the
following evolution equations%
\begin{equation}
\stackrel{.}{c}_{d}=-\sqrt{\kappa /2\pi }\sin \theta \int_{-\omega
_{b}}^{+\omega _{b}}c_{\omega }d\omega ,  \label{10}
\end{equation}%
\begin{equation}
\stackrel{.}{c}_{\omega }=-i\omega c_{\omega }+\sqrt{\kappa /2\pi }c_{d}\sin
\theta .  \label{11}
\end{equation}%
Equation (13) has the solution%
\begin{equation}
c_{\omega }\left( t\right) =\sqrt{\kappa /2\pi }\int_{0}^{t}e^{-i\omega
\left( t-\tau \right) }c_{d}\left( \tau \right) \sin \theta \left( \tau
\right) d\tau ,  \label{12e}
\end{equation}%
which, substituted into Eq. (12), leads to%
\begin{eqnarray}
\stackrel{.}{c}_{d} &=&-\frac{\kappa }{2\pi }\sin \theta \int_{0}^{t}\frac{%
\sin \left[ \omega _{b}\left( t-\tau \right) \right] }{t-\tau }c_{d}\left(
\tau \right) \sin \theta \left( \tau \right) d\tau   \nonumber \\
&\simeq &-\left( \kappa /2\right) c_{d}\sin ^{2}\theta .  \label{13}
\end{eqnarray}%
The approximation in Eq. (15) is valid since the bandwidth $\omega _{b}$
satisfies $\omega _{b}T\gg 1$, where the operation time $T$ characterizes
the time scale for a significant change of $c_{d}$ and $\sin \theta $.
Therefore, the dark-state coefficient $c_{d}$ satisfies the cavity
free-decay equation, with the decay rate $\kappa $ replaced by the effective
rate $\kappa \sin ^{2}\theta $. This can be easily understood since $\sin
^{2}\theta $ is the probability of the component $\left| s\right\rangle
\left| 1\right\rangle $ in the dark state $\left| D\right\rangle $, and it
is exactly this component that couples to the cavity output. Equation (15)
has the straightforward solution%
\begin{equation}
c_{d}=\exp \left( -\frac{\kappa }{2}\int_{0}^{t}\sin ^{2}\theta \left( \tau
\right) d\tau \right) .  \label{14e}
\end{equation}%
We want to know the single-photon pulse shape $f\left( t\right) $ of the
cavity output state $\left| \varphi _{1}\right\rangle $. Suppose now that $T$
is the final time of the interaction (i.e., the operation time determined by
the driving laser pulse is from $0$ to $T$). The pulse shape $f\left(
t\right) $ is connected with the coefficients $c_{\omega }\left( t\right) $
before the frequency components in $\left| \varphi _{1}\right\rangle $ by
the Fourier transformation \cite{19}%
\begin{equation}
f\left( t\right) =\frac{1}{\sqrt{2\pi }}\int_{-\omega _{b}}^{+\omega
_{b}}d\omega c_{\omega }\left( T\right) e^{-i\omega \left( t-T\right) }.
\label{15}
\end{equation}%
From Eqs. (14), (16), and (17), we finally obtain%
\begin{equation}
f\left( t\right) =\sqrt{\kappa }\sin \theta \left( t\right) \exp \left( -%
\frac{\kappa }{2}\int_{0}^{t}\sin ^{2}\theta \left( \tau \right) d\tau
\right) .  \label{16}
\end{equation}%
Note that the single-photon pulse shape $f\left( t\right) $ is completely
determined by $\theta \left( t\right) $, i.e., by the driving pulse shape $%
\widetilde{\varepsilon }\left( t\right) $, and is independent of the random
atom position ${\bf r}$ and the absolute value of the coupling coefficient $%
g\left( {\bf r}\right) $. As we have mentioned before, this is the main
advantage of this adiabatic method compared with either the Raman scheme or
prior proposals based upon adiabatic passages with uniform illumination \cite%
{15',16,17}, and this feature is essential for many applications of this
setup.

The above result is obtained within the adiabatic approximation, and in the
adiabatic limit, the solution is independent of the atomic spontaneous
emission rate $\gamma _{s}$ and the detuning $\Delta $. This is only a rough
picture. In the following, we will solve exactly the dynamics governed by
the Hamiltonian (6) without the use of the adiabatic approximation. The
exact solution is necessary in the following two senses: first, we need to
verify the above ideal picture and to find out under what condition this
picture is approximately valid. Though in the three-level case, we have some
simple estimation of the condition for the adiabatic following, it is not
easy to figure out the exact adiabatic following condition for more
realistic situation of a continuum of external modes. In this case, the
argument based on the level spacing is not valid. We need to know how long
the operation time $T$ should be to satisfy the adiabatic following
condition. We also expect that the atomic spontaneous emission can not be
made negligible simply by increasing the operation time $T$. Its rate $%
\gamma _{s}$ should be small enough to satisfy the strong coupling condition
$\kappa \gamma _{s}\ll \overline{g}^{2}$, where $\overline{g}$ denotes the
average of the coupling rate $g\left( {\bf r}\right) $ \cite{19'}. Second,
in real experiments, the operation time $T$ is not infinitely long, and the
coupling rate $\overline{g}$ can not be arbitrarily larger than the decay
rates $\kappa $ and $\gamma _{s}$ due to limitation of the technology (for
instance, in Caltech experiments, typically, $\overline{g}/2\pi $ is around $%
20$MHz, and $\kappa /2\pi \sim \gamma _{s}/2\pi \sim 6$MHz). In this case,
there would be various non-adiabatic corrections to the above ideal picture,
for instance, the atom may go down from the level $\left| e\right\rangle $
to $\left| s\right\rangle $ through a spontaneous emission, and then we lose
the emitted photon and thus have no output from the cavity; or we have a
single-photon output, but it is in a wrong and unknown pulse shape due to
its sensitivity to the random atom position induced by the non-adiabatic
contributions. It is desirous to calculate quantitatively the magnitudes of
these noises to predict the real experiments. The exact solution of the
system dynamics is only available with the numerical methods, which is the
main task of the following section.

\section{Exact numerical simulations}

\subsection{The numerical calculation method}

In this section, we solve exactly the system dynamics governed by the
Hamiltonian (6) through numerical simulations, and calculate various
nonadiabatic corrections and noise magnitudes. For numerical simulations of
the Hamiltonian (6), we need to discretize the free-space filed $b\left(
\omega \right) $ by introducing a finite but small frequency interval $%
\delta \omega $ between two adjacent modes. Then, in total we have about $%
N\approx 2\omega _{b}/\delta \omega $ free-space modes, with the $j$ mode
denoted by $b_{j}$. The frequency detuning $\omega _{j}$ of the $j$ mode is
given by $\omega _{j}=(j-N/2)\delta \omega $. To assure that there is no
change of the physical result after the discretization, we should choose the
frequency interval $\delta \omega $ much smaller than the inverse of the
operation time $T$, and the bandwidth $\omega _{b}$ much larger the cavity
decay rate $\kappa $.

For the numerical simulation, we can similarly expand the state $\left| \Psi
\right\rangle $ of the whole system in the form of Eq. (9), with the
single-photon pulse state replaced by
\begin{equation}
\left| \varphi _{1}\right\rangle =\sum_{j=1}^{N}c_{j}b_{j}^{\dagger }\left|
\text{vac}\right\rangle .  \label{phi1again}
\end{equation}%
From the Hamiltonian (6), we get the following complete set of equations for
the coefficients $c_{d},$ $c_{b},$ $c_{e},$ and $c_{j}$%
\begin{equation}
\stackrel{.}{c}_{d}=-\stackrel{.}{\theta }c_{b}-\kappa ^{\prime }\sin \theta
\sum_{j=1}^{N}c_{j},  \label{17}
\end{equation}%
\begin{equation}
\stackrel{.}{c}_{b}=\stackrel{.}{\theta }c_{d}-i\sqrt{\Omega ^{2}\left( {\bf %
r},t\right) +g^{2}\left( {\bf r}\right) }c_{e}+\kappa ^{\prime }\cos \theta
\sum_{j=1}^{N}c_{j},  \label{18}
\end{equation}%
\begin{equation}
\stackrel{.}{c}_{e}=\left( -i\Delta -\gamma _{s}/2\right) c_{e}-i\sqrt{%
\Omega ^{2}\left( {\bf r},t\right) +g^{2}\left( {\bf r}\right) }c_{b},
\label{19}
\end{equation}%
\begin{equation}
\stackrel{.}{c}_{j}=-i\left( j-N/2\right) \delta \omega c_{j}+\kappa
^{\prime }\sin \theta c_{d}-\kappa ^{\prime }\cos \theta c_{b},  \label{20}
\end{equation}%
where the effective decay rate $\kappa ^{\prime }\equiv \sqrt{\kappa \delta
\omega /2\pi }$. We obtain the solutions of the these coefficients by
numerically integrating Eqs. (20)-(23) from the time $t=0$ to $t=T$, where $T
$ is the duration of the driving pulse $\widetilde{\varepsilon }\left(
t\right) $. We assume that $\widetilde{\varepsilon }\left( t\right) $ is \ a
Gaussian pulse so that $\alpha \left( t\right) $ is a Gaussian function of
the time $t$, with its peak value at $T/2$, and a width $t_{w}$
significantly smaller than $T/2$. All the functions of $\theta $\ in Eqs.
(20)-(23) are decided from $\cos \theta =1/\sqrt{1+\left| r_{o}\alpha \left(
t\right) \right| ^{2}}$, and $\sqrt{\Omega ^{2}\left( {\bf r},t\right)
+g^{2}\left( {\bf r}\right) }=g\left( {\bf r}\right) /\cos \theta $. To
simulate the randomness of the atom position ${\bf r}$, we vary the value of
$g\left( {\bf r}\right) $ in the simulation to look at whether the final
result changes with this variation.

\subsection{Shape of the output single-photon pulse}

The output single-photon pulse shape $f\left( t\right) $ can be easily
constructed from the solution of the coefficients $c_{j}$ through a discrete
version of Eq. (17). The result is shown in Fig. 2 for $g\left( {\bf r}%
\right) =3\kappa $ and $g\left( {\bf r}\right) =6\kappa $. Although we have
not made definitive measurements, we estimate that $g\left( {\bf r}\right) $
varies within a factor of roughly $2$ in the current Caltech experiment\cite%
{2,icols01,mckeever02}. Here and in the following, the pulse shape function $%
f\left( t\right) $ is always re-normalized according to $\int \left| f\left(
t\right) \right| ^{2}dt=1$ for convenience of comparison. We see that the
two curves overlap very well, which confirms the prediction that the output
pulse shape is very insensitive to the randomness of the coupling
coefficient $g\left( {\bf r}\right) $ when the adiabatic condition is
satisfied (we take $T=20/\kappa $ for this figure). We also draw in this
figure the pulse shape $f\left( t\right) $ given by Eq. (18) derived in the
ideal adiabatic limit, which agrees well with the exact numerical results.
Therefore, within the adiabatic condition, we can use the analytical result
(18) to design the shape of the output single-photon pulse by modulating the
driving pulse shape $\widetilde{\varepsilon }\left( t\right) $.

\begin{figure}[tbp]
\epsfig{file=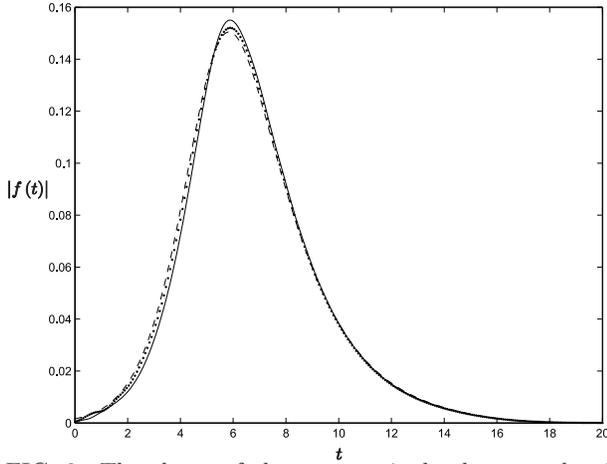,width=8cm} \caption{The shape of the output
single-photon pulse described by the amplitude $\left| f\left(
t\right) \right| $ versus the time $t$ for the
coupling rates $g\left( {\bf r}\right) =3\protect\kappa $ (solid curve) and $%
g\left( {\bf r}\right) =6\protect\kappa $ (dotted curve). The dashed curve
represents the pulse shape in the ideal adiabatic limit calculated from Eq.
(16). In this figure, we have taken $\protect\gamma _{s}=\protect\kappa $, $%
\Delta =0$, and $T=20/\protect\kappa $. The driving pulse $\widetilde{%
\protect\varepsilon }\left( t\right) $ is in a Gaussian shape with the peak
at $t=T/2$ and a width $t_{w}=T/5$.}
\end{figure}

\subsection{Noise magnitudes and the adiabatic condition}

To quantify the noise magnitudes in this setup, we can define several error
probabilities. First, we have the leakage error due to the atomic
spontaneous emission. A photon may be emitted to modes other than the
principal cavity mode through the spontaneous emission with the rate $\gamma
_{s}$. As a result, the norm $\left| c_{d}\right| ^{2}+\left| c_{b}\right|
^{2}+\left| c_{e}\right| ^{2}+\sum_{j=1}^{N}\left| c_{j}\right| ^{2}$ of the
state (10) decays with the time $t$, and we can use
\begin{equation}
P_{\text{spon}}=1-\left| c_{d}\left( T\right) \right| ^{2}-\left|
c_{b}\left( T\right) \right| ^{2}-\left| c_{e}\left( T\right) \right|
^{2}-\sum_{j=1}^{N}\left| c_{j}\left( T\right) \right| ^{2}  \label{21}
\end{equation}%
at the final time $T$ to quantify the total possibility of the spontaneous
emission loss. Second, due to the finiteness of the operation time $T$ and
the pumping field amplitude $\widetilde{\varepsilon }\left( t\right) $, the
initial excitation in the dark state is not necessarily fully transferred to
the output quantum signal at the final time, and we can use%
\begin{equation}
P_{\text{tran}}=\left| c_{d}\left( T\right) \right| ^{2}+\left| c_{b}\left(
T\right) \right| ^{2}+\left| c_{e}\left( T\right) \right| ^{2}  \label{22}
\end{equation}%
at the time $T$ to quantify the transmission inefficiency. In principle, we
can arbitrarily decrease the transmission inefficiency by increasing the
duration $T$ or the amplitude $\widetilde{\varepsilon }\left( t\right) $ of
the pumping field. Finally, even if a photon is emitted into the cavity
output field, it is not necessarily in the right pulse shape as given by Eq.
(18) due to the non-adiabatic correction. This non-adiabatic correction
depends on the random atom position and is unknown, so it is also a source
of noise. To quantify this noise, we denote the ideal pulse shape given in
Eq. (18) as $f_{\text{id}}\left( t\right) $, and the real pulse shape
calculated from the numerical simulation as $f_{\text{real}}\left( t\right) $%
, then the shape mismatching error can be described by
\begin{equation}
P_{\text{mis}}=\left| 1-\frac{\int_{0}^{T}f_{\text{real}}^{\ast }\left(
t\right) f_{\text{id}}\left( t\right) dt}{\left[ \int_{0}^{T}\left| f_{\text{%
real}}\left( t\right) \right| ^{2}dt\int_{0}^{T}\left| f_{\text{id}}\left(
t\right) \right| ^{2}dt\right] ^{1/2}}\right| .  \label{23}
\end{equation}%
This quantity is directly related to the visibility of the fringes if we
interfere two single-photon pulses from two such setups.

For the example shown in Fig. 2, with $g\left( {\bf r}\right) =3\kappa
=3\gamma _{s}$ (the other parameters are given in the figure caption), we
have $P_{\text{spon}}\approx 4.0\%$, $P_{\text{tran}}\approx 0.04\%$, $P_{%
\text{mis}}\approx 0.18\%$. The dominant source of noise is the leakage
error $P_{\text{spon}}$ induced by the spontaneous emission. If we increase
the operation time $T$ so that the adiabatic condition is better satisfied,
the above-defined noise magnitudes can be reduced a little bit, but not too
much. For instance, with the above example but $T=30/\kappa $, we have $P_{%
\text{spon}}\approx 3.33\%$ and $P_{\text{mis}}\approx 0.15\%$. On the other
hand, if $T$ is reduced so that the adiabatic condition is not well
satisfied, the error probabilities can significantly increase. Fig. 3 shows
the output pulse shapes for $g\left( {\bf r}\right) =3\kappa $ and $g\left(
{\bf r}\right) =6\kappa $ with $T=5/\kappa $. The two curves are obviously
different to each other and are also different to the ideal shape as given
by Eq. (18). For the example with $g\left( {\bf r}\right) =3\kappa =3\gamma
_{s}$ and $T=5/\kappa $, we have $P_{\text{spon}}\approx 36\%$, $P_{\text{%
tran}}\approx 3.2\%$, $P_{\text{mis}}\approx 2.7\%$. All the noise
magnitudes significantly increase. In particular, the spontaneous emission
loss becomes very big. This can be easily understood since without the
adiabatic condition, the excited state $\left| e\right\rangle $ will be
populated during the operation, and thus we have a correspondingly larger
spontaneous emission loss.

\begin{figure}[tbp]
\epsfig{file=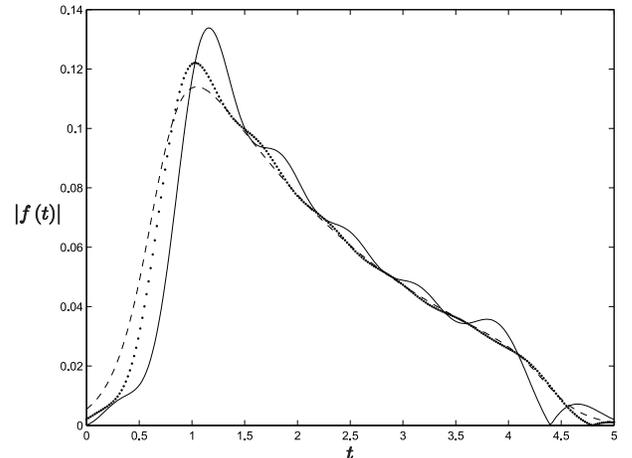,width=8cm} \caption{The shape $\left|
f\left( t\right) \right| $ of the output
single-photon pulse for the coupling rates $g\left( {\bf r}\right) =3\protect%
\kappa $ (solid curve), $g\left( {\bf r}\right) =6\protect\kappa $
(dotted curve), and in the ideal adiabatic limit (dashed curve).
We assumed the same condition as in Fig. 2, except that
$T=5/\protect\kappa $, which does not well satisfy the adiabatic
condition.}
\end{figure}

\subsection{The strong coupling condition}

Next we look at the requirement of the strong coupling condition. Let $%
\overline{g}$ denote the average value of the coupling rate $g\left( {\bf r}%
\right) $. Normally one requires $\overline{g}^{2}\gg \kappa \gamma _{s}$ to
satisfy the strong coupling condition. We can define the strong coupling
parameter $d_{sc}$ as $d_{sc}=\overline{g}^{2}/\kappa \gamma _{s}$, and
calculate the above-defined noise magnitudes $P_{\text{spon}}$,$P_{\text{tran%
}}$,$P_{\text{mis}}$ under different values of the parameter $d_{sc}$. We
assumed $T=30/\kappa $ and $\Delta =0$ in the calculation so that the
adiabatic condition is well satisfied. It turns out that the spontaneous
emission loss $P_{\text{spon}}$ is always the dominant loss (about ten times
larger than other sources of noise). Thus, in Fig. 4, we only show the
calculation result for $P_{\text{spon}}$ under different values of $d_{sc}$.
The result can be approximately simulated by an empirical curve with $P_{%
\text{spon}}\approx 1/\left( 4d_{sc}\right) $.

We can use this simple formula to estimate the spontaneous emission loss
under different experimental conditions. Actually, in current experiments,
the strong coupling condition is only marginally satisfied. For instance,
for the cesium atom in the Caltech group, $\left( \kappa ,\gamma _{s}\right)
/2\pi \approx \left( 8,5.2\right) $ MHz (note that $\kappa $ and $\gamma _{s}
$ here denote the energy decay rates, which are two times the corresponding
amplitude decay rates) \cite{2,icols01}, and $\overline{g}/2\pi $ is
expected to be approximately $15$ MHz for the transition $\left(
6S_{1/2},F=4,m=+4\right) \longrightarrow \left( 6P_{3/2},F=4,m=+4\right) $
(Note that the transition $\left( 6S_{1/2},F=4,m=+4\right) \longrightarrow
\left( 6P_{3/2},F=5,m=+5\right) $ cannot be used as a $\Lambda -$%
configuration though it has a slightly larger coupling rate $\overline{g}$).
These values lead to $d_{sc}=\overline{g}^{2}/\kappa \gamma _{s}\approx 5.4$
and a resulting spontaneous emission loss around $4.6\%,$ which is quite
accessible with the present technology. As another example, in the recent
experiment \cite{17}, one has $\left( \kappa ,\gamma _{s}\right) /2\pi
\approx \left( 1.25,6.0\right) $ MHz, and $\overline{g}/2\pi \approx 2.5$
MHz  according to the estimation there. With these parameters, we estimate
that the spontaneous emission loss is about $P_{\text{spon}}\approx 30\%$ if
one uses the scheme here. If the usual adiabatic scheme is adopted with a
uniform driving pulse perpendicular to the cavity axis, the spontaneous
emission loss should be still significantly larger, as will be seen from the
simulation in the last subsection.

\begin{figure}[tbp]
\epsfig{file=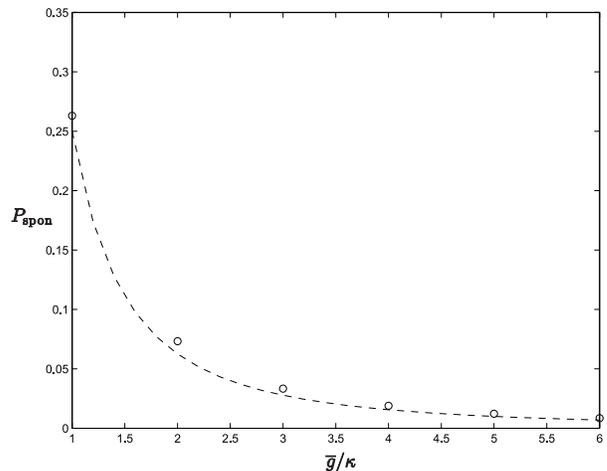,width=8cm} \caption{The spontaneous
emission loss $P_{\text{spon}}$ versus the average
coupling rate $\overline{g}$(in the unit of the cavity decay rate $\protect%
\kappa $). We assumed $\Delta=0 $ and $\protect\gamma _{s}= \protect\kappa $%
, so the strong coupling parameter $d_{sc}$ is simply $({\overline{g}/%
\protect\kappa})^2 $. The circles represent the results from the numerical
calculation, and the dashed curve is from the empirical formula $P_{\text{%
spon}}\approx 1/\left( 4d_{sc}\right) $ which well simulates the numerical
results. }
\end{figure}

\subsection{The influence of the single-photon transition detuning}

In the above calculations, we assumed $\Delta =0$. Finally, we discuss the
influence of a non-zero single-photon detuning $\Delta $. In Fig. 5, we show
the calculation result of the exact pulse shape function $f_{\text{real}%
}\left( t\right) $ with a significant detuning $\Delta =\kappa $, and
compare it with the ideal pulse shape function $f_{\text{id}}\left( t\right)
$ given by Eq. (18) for both of the amplitude and the phase. The other
parameters for this example are given in the figure caption. From the
figure, we see that the two amplitudes $\left| f_{\text{real}}\left(
t\right) \right| $ and $\left| f_{\text{id}}\left( t\right) \right| $ still
overlap very well, but their phases become a bit different due to the
detuning.

This phase difference is determined by the the detuning $\Delta $, whereas
the latter depends on the different level shift between ground and excited
states, and hence varies with atom position within the FORT beam. In the
case of the simple level scheme depicted in Figure 1, the states $\left|
g\right\rangle $ and $\left| e\right\rangle $ would have spatially dependent
level shifts of opposite sign, which would lead to variations in $\Delta $
comparable to the trap depth. Fortunately, there is a simple avenue to
mitigate this difficulty by considering the multi-levels involved for the
FORT beam, as described in Ref. \cite{icols99}, so that the trapping
potentials for the states $\left| g\right\rangle $ and $\left|
e\right\rangle $ are very nearly the same. For example, for the experiment
of Ref. \cite{mckeever02}, the difference in trap depth for $\left|
g\right\rangle $ and $\left| e\right\rangle $ is roughly $10$\% of the trap
depth. Relative to the current analysis, there is then a variation in $%
\Delta $ as the atom moves in the FORT\ potential, which is unknown when the
adiabatic protocol is implemented. The curve in Fig. 5 is an attempt to
estimate the impact of such random detunings by setting $\Delta =\kappa $,
which exceeds the actual magnitude of any spatially dependent detunings for
FORT depths up to about $50$MHz. The phase difference in the pulse shape
function caused by the unknown detunings is a source of noise, which
contributes to the shape mismatching error defined in Eq. (26). For this
example with $g\left( {\bf r}\right) =3\kappa $, we have $P_{\text{spon}%
}\approx 3.33\%$, $P_{\text{tran}}\approx 10^{-4}$, which are basically the
same to the corresponding case without detuning, but $P_{\text{mis}}\approx
3.33\%$, which becomes significantly larger due to the contribution of the
phase difference.

\begin{figure}[tbp]
\epsfig{file=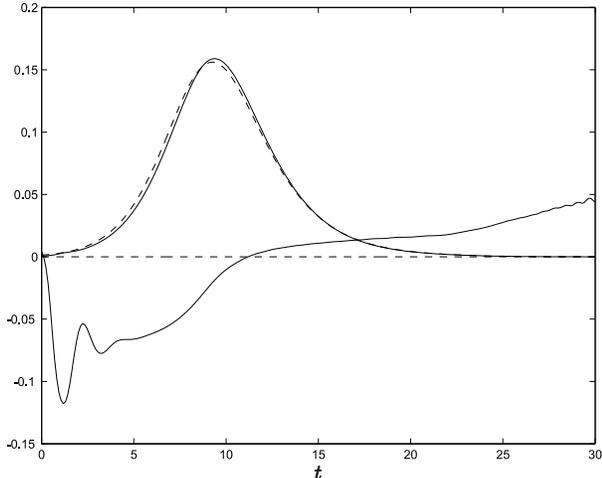,width=8cm}
\caption{The amplitude (the modular) and the phase (divided by $\protect\pi %
/2$) of the real pulse shape $f_{\text{real}}\left( t\right) $ (two solid
curves) and the ideal pulse shape $f_{\text{id}}\left( t\right) $ (two
dashed curves) versus the time $t$ with the single-photon transition
detuning $\Delta =\protect\kappa $. We assumed $g\left( {\bf r}\right) =3%
\protect\kappa $, $\protect\gamma _{s}=\protect\kappa $, and $T=30/\protect%
\kappa $. In this case, the main difference between $f_{\text{real}}\left(
t\right) $ and $f_{\text{id}}\left( t\right) $ lies in the phase difference.}
\end{figure}

\subsection{Compare with the usual adiabatic scheme}

In our scheme, the driving pulse is matched to a cavity mode which has
basically the same spatial mode structure as the cavity QED light. In usual
adiabatic schemes \cite{15',17}, the driving laser is assumed to be
perpendicular to the cavity axis with uniform illumination intensity. We
expect that with the present interaction configuration, our scheme is more
insensitive to the randomness in the atom position. To compare the two
configurations more quantitatively, we have calculated the output pulse
shapes and noise magnitudes for both schemes.

First, let us assume that the atom has been trapped in one potential well,
but the coupling rate $g\left( {\bf r}\right) $ may vary within a factor $2$
due to the unknown atom position. In figure 6, we show the calculation
results of the output pulse shapes. The solid curve shows the pulse shape
function $\left| f\left( t\right) \right| $ when $g\left( {\bf r}\right)
=3\kappa $ and $\Omega _{m}\left( {\bf r}\right) =3\kappa $, where $\Omega
_{m}\left( {\bf r}\right) $ is the maximum of $\Omega \left( {\bf r}%
,t\right) $ with respect to time $t$ ($\Omega \left( {\bf r},t\right) $ is
assumed to be a Gaussian function of $t$ as specified in the caption of Fig.
2). Now, if $g\left( {\bf r}\right) $ varies by a factor $2$ due to change
of the atom position, in our scheme the Rabi frequency will correspondingly
change by the same ratio. The dashed curve shows the pulse shape for $%
g\left( {\bf r}\right) =6\kappa $ and $\Omega _{m}\left( {\bf r}\right)
=6\kappa $. One can see that the two curves overlap very well with the mode
mismatching noise smaller than $0.2\%$. In contrast, in usual adiabatic
schemes with uniform illumination intensity, $\Omega _{m}\left( {\bf r}%
\right) $ does not change as $g\left( {\bf r}\right) $ varies with the atom
position, so we have the same $\Omega _{m}\left( {\bf r}\right) =3\kappa $.
The dotted curve in Fig. 6 shows the pulse shape for $g\left( {\bf r}\right)
=6\kappa $ and $\Omega _{m}\left( {\bf r}\right) =3\kappa $. It is
significantly different to the above two curves with a notable mode
mismatching noise $P_{\text{mis}}\approx 6.9\%$. Therefore, by this
interaction configuration, the scheme is more robust to the random variation
of the atom position.

\begin{figure}[tbp]
\epsfig{file=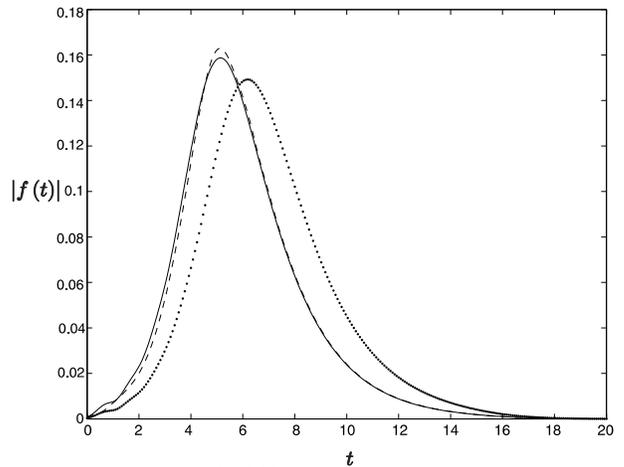,width=8cm} \caption{The shape $\left|
f\left( t\right) \right| $ of the output single-photon pulse for
the following pairs of coupling rates and the driving Rabi
frequencies: first, $g\left( {\bf r}\right) =3\protect%
\kappa $ and $\Omega _{m}\left( {\bf r}\right) =3\kappa $ (solid
curve), second, $g\left( {\bf r}\right) =6\protect\kappa $ and
$\Omega _{m}\left( {\bf r}\right) =6\kappa $(dashed curve), and
finally, $g\left( {\bf r}\right) =6\protect\kappa $ and $\Omega
_{m}\left( {\bf r}\right) =3\kappa $ (dotted curve). The other
parameters are the same as in Fig. 2.}
\end{figure}

The improvement by this protocol is more remarkable if we consider the case
where the atom is not fixed in one potential well, and may jump form well to
well in the axial direction, as is the case in some recent experiments \cite%
{16,17}. The variation of the atom position in the axial direction is
typically fast compared with the operation time $T$, so we have a
time-varying atom position ${\bf r}$ and coupling rate $g\left( {\bf r}%
\right) $. Here, we consider an explicit form of the time variation of $%
g\left( {\bf r}\right) $ by assuming $g\left( {\bf r}\right) =6\kappa \sin
\left( 4\pi t/T+\varphi _{0}\right) $, where the phase $\varphi _{0}$ is
randomly chosen corresponding to the randomness in the initial atom
position. It is enough to illustrate the general result by considering this
special example. First, let us calculate the output pulse shape $f\left(
t\right) $ for the usual adiabatic scheme where $\Omega _{m}\left( {\bf r}%
\right) $ is fixed as a constant \cite{15',16,17}. The solid and the dashdot
curves in Fig. 7 show the real parts of $f\left( t\right) $ with initial
phase $\varphi _{0}=0$ and $\varphi _{0}=\pi /2$, respectively (the
imaginary parts of $f\left( t\right) $ are actually small and negligible).
The two curves do not overlap at all. Neither the magnitude nor the phase of
the pulse shape $f\left( t\right) $ can be controlled with this scheme. We
also calculate the spontaneous emission loss $P_{\text{spon}}$ for this
example. The average spontaneous emission loss is about $P_{\text{spon}%
}\approx 25\%$.

\begin{figure}[tbp]
\epsfig{file=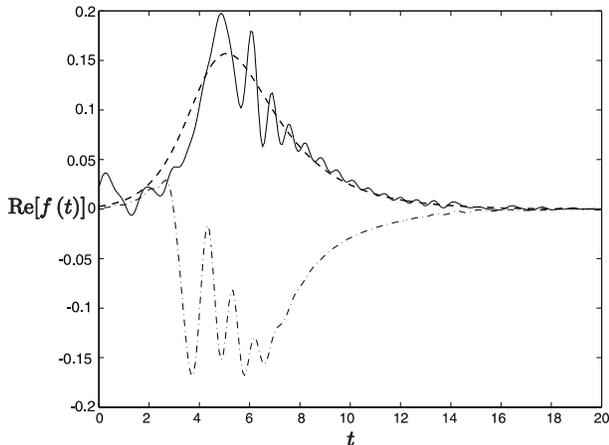,width=8cm} \caption{The real part of the
pulse shape function $Re\left[ f\left( t\right) \right] $ as
$g\left( {\bf r}\right)$ varies with time by the form $g\left(
{\bf r}\right) =6\kappa \sin \left( 4\pi t/T+\varphi _{0}\right) $
in the usual adiabatic scheme with $\varphi _{0}=0$ (solid curve)
and $\varphi _{0}=\pi /2$ (dashdot curve), respectively. The
dashed curve shows the ideal pulse shape calculated from Eq. (18).
The other parameters in this figure are the same as those in Fig.
2.}
\end{figure}

Similarly, we can calculate the pulse shape for the same example with the
present scheme. In this case, due to the atomic motion, $\Omega _{m}\left(
{\bf r}\right) $ varies with time in the same way as $g\left( {\bf r}\right)
$, but the ratio $\Omega _{m}\left( {\bf r}\right) /g\left( {\bf r}\right) $
is kept constant. Fig. 8 shows the real part of the shape function $f\left(
t\right) $ in this case, with the solid and the dashdot curves corresponding
to the initial phase $\varphi _{0}=0$ and $\varphi _{0}=\pi /2$,
respectively. Although the two curves do not overlap very well, they still
look similar with the same phase. They also roughly agree with the ideal
shape function given by Eq. (18), which is shown as the dashed curve in Fig.
8. The average mode mismatching noise for these two curves is given by $P_{%
\text{mis}}\approx 1.1\%$, and the average spontaneous emission loss is $P_{%
\text{spon}}\approx 9.4\%$. The spontaneous emission loss is also
significantly reduced with the present scheme. This can be understood as
follows: if one has a constant $\Omega _{m}\left( {\bf r}\right) $ as the
usual adiabatic scheme, when the atom moves to the place with $g\left( {\bf r%
}\right) $ near to zero, the adiabatic condition is not well satisfied, and
as a result, one has a considerably large spontaneous emission loss;
however, in the present scheme, in the place where $g\left( {\bf r}\right) $
is near to zero, $\Omega _{m}\left( {\bf r}\right) $ is also near to zero.
The excitation probability of the atom is then reduced, and the adiabatic
condition is better satisfied. Consequently, one has a smaller spontaneous
emission loss.

\begin{figure}[tbp]
\epsfig{file=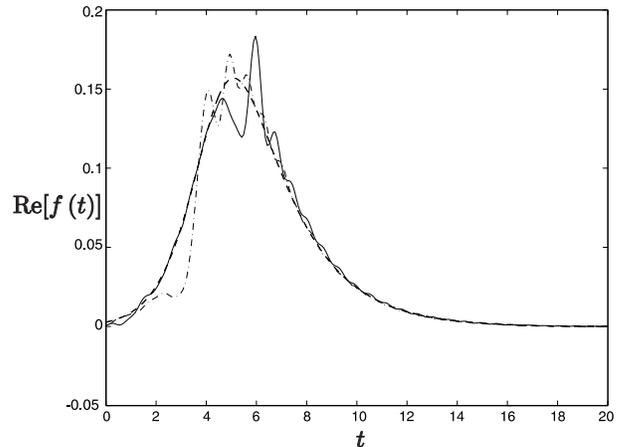,width=8cm} \caption{The real part of the
pulse shape function $Re\left[ f\left( t\right) \right] $
calculated for the same example as in Fig.7, but now for the
present adiabatic scheme where the driving Rabi frequency varies
in the same way as the the coupling rate when the atom moves.}
\end{figure}

\section{Applications}

There have been many proposals to use the setup with single atoms in high-Q
cavities for various applications, such as for single-photon or entangled
photon source \cite{5,6}, for quantum communication between different
cavities \cite{8}, for atomic quantum teleportation \cite{9,10}, and for
quantum computation \cite{11}. In these proposals, one always assumed the
atom is well localized so that the Lamb-Dick condition is satisfied.
However, one can apply the method here to all of the schemes mentioned above
to eliminate \ the challenging Lamb-Dick condition. Basically, what one
needs to do is to replace the Raman scheme with the adiabatic scheme, and to
keep the pumping laser collinear with the cavity axis so that the driving
pulse and the cavity mode have the same spatial mode structure. All the
calculation results (for the noise magnitudes, pulse shape, etc.) in this
paper apply to these schemes. After the improvement, it becomes considerably
easier to implement these schemes with the current technology. Here, we
briefly mention some possibilities.

\subsection{Controllable single-photon or entangled-photon source}

It is desirable to have a single-photon source with all its properties fully
controllable, including its emission direction, emission time, and pulse
shape. This kind of source has important application in some recent quantum
information processing schemes \cite{20}, which are normally based on the
interference of different single-photon pulses. To get interference between
deferent single-photon pulses, one needs to require all the pulses to be
directional and to have the same time shape. Recently, there have been
significant experimental advances in the realization of the single-photon
source \cite{21,22,23,16,17}. In the experiments based on the solid-state
material \cite{21,22,23}, the single-photon emitter has a fixed position,
and one can in principle well control the pulse shape. However, the emitted
pulse is typically not directional. On the other hand, in current
experiments \cite{16,17} with high-Q cavities, the emitted pulse is
directional, but its shape is not well controlled since with uniform
illumination of a perpendicular driving pulse, the waveform $f(t)$ depends
on the time history of the coupling rate $g\left( {\bf r}\right) $, which in
turn depends on the atom's position. As the atom falls through the cavity,
it has basically a random trajectory, leading to unknown variations in $%
g\left( {\bf r}\right) $ both in magnitude and sign. It is a challenging
experimental endeavor to demonstrate a single-photon source with all the
properties mentioned above fully controllable.

The method in this paper shows that the single atom trapped in a high-Q
cavity is a good candidate for the realization of the fully controllable
single-photon source. Though the coupling rate $g\left( {\bf r}\right) $ is
not completely fixed in current setups due to the hardness to fully localize
the atom, the emitted single-photon pulse has a definitely well controllable
time shape and emission direction with an appropriate design of the
interaction configuration as has been shown before.

As shown in Ref. \cite{6}, with a more involved atomic level structure, it
is possible to engineer entanglement between different single-photon pulses.
It is straightforward to combine the method here with that scheme to
eliminate the requirement of the Lamb-Dick condition in \cite{6} so that one
can get an entangled single-photon source with the ``hot'' trapped atom as
well.

\subsection{Quantum communication between different cavities}

The dynamics governed by the Hamiltonian (6) is reversible if we neglect the
atomic spontaneous emission $\gamma _{s}$. Therefore, if one directs the
emitted single-photon pulse back to the cavity, and at the same time reverse
both of the time shapes of the single-photon pulse and the driving pulse,
the single-photon pulse will be completely absorbed as long as the noise
effects are negligible. It was first proposed in \cite{8} that one can use
this kind of phenomenon to achieve quantum communication between different
cavities, that is, to transfer quantum states of a trapped atom from one
cavity to another cavity. For this purpose, one can require that the emitted
single-photon pulse has a time-symmetric shape by modulating the driving
pulse shape. For a time-symmetric pulse, its time reversal is itself, so we
can directly direct this pulse to another cavity with the same configuration
but with a time-reversed driving pulse, then the single-photon pulse will be
completely absorbed by this cavity, which transfers the atomic state from
one cavity to the other one. The scheme in \cite{8} is based on the Raman
configuration, but it is straightforward to transfer it to the adiabatic
configuration discussed in this paper so that it works with a hot trapped
atom. Note that the same setup can also be used for storage of a
single-photon pulse with a known shape \cite{18,24,25}.

To get a time-symmetric single-photon pulse for a complete absorption of the
second cavity, Ref. \cite{8} gives a numerical solution to the shape of the
driving pulse. For the adiabatic configuration, within a good approximation,
we have an analytic expression (18) which connects the shape of the output
single-photon pulse to the shape of the driving pulse. With this analytical
expression, it becomes easier to design the shape $\widetilde{\varepsilon }%
\left( t\right) $ of the driving pulse. For this purpose, we solve $\sin
\theta \left( t\right) $ from Eq. (18) as%
\begin{equation}
\sqrt{\kappa }\sin \theta \left( t\right) =\frac{f\left( t\right) }{\sqrt{%
1-\int_{0}^{t}f^{2}\left( \tau \right) d\tau }}.  \label{24}
\end{equation}%
The form of $\sin \theta \left( t\right) $ is immediately available from
this equation for any desirable output pulse shape $f\left( t\right) $
(which has been assumed to be real and positive for simplicity). Then, the
shape of the driving pulse can be easily decided from $\widetilde{%
\varepsilon }\left( t\right) \propto \alpha \left( t\right) $ and $\sin
\theta \left( t\right) =r_{o}\alpha \left( t\right) /\sqrt{1+\left|
r_{o}\alpha \left( t\right) \right| ^{2}}$, where $r_{o}$ is the ratio of
the Clebsch-Gordan coefficients. For instance, if we want to have a time
symmetric $f\left( t\right) $ in the period $0\leq t\leq T$ with the form $%
f\left( t\right) =\sqrt{\beta /2}%
\mathop{\rm sech}%
\left[ \beta \left( t-T/2\right) \right] $ , where we have assumed $%
\mathop{\rm sech}%
\left( -\beta T/2\right) \ll 1$, $\sin \theta \left( t\right) $ should be in
the form $\sin \theta \left( t\right) =\sqrt{\beta /\kappa }\sqrt{1+\tanh %
\left[ \beta \left( t-T/2\right) \right] }$. Note that we only have a
solution of $\theta \left( t\right) $ when the rate $\beta <\kappa /2$,
which is consistent with the observation that any pulse from the decay of a
cavity can not vary with the time faster than the cavity decay rate. From $%
\sin \theta \left( t\right) $, we see that the shape $\widetilde{\varepsilon
}\left( t\right) $ of the driving pulse should be chosen according to%
\begin{equation}
\widetilde{\varepsilon }\left( t\right) \propto \sqrt{\frac{1+\tanh \left[
\beta \left( t-T/2\right) \right] }{\left( \kappa /\beta -1\right) -\tanh %
\left[ \beta \left( t-T/2\right) \right] }}.  \label{25}
\end{equation}%
As a special case, if $\kappa /\beta =2$, $\widetilde{\varepsilon }\left(
t\right) \propto e^{\beta \left( t-T/2\right) }$, which grows exponentially
with the time $t$ for the operation period $0\leq t\leq T$. Therefore, we
have\ a simple solution to the driving pulse shape for quantum communication
between two different cavities: for the first cavity, we apply an
exponentially increasing pulse with $\widetilde{\varepsilon }\left( t\right)
=\widetilde{\varepsilon }\left( 0\right) e^{\kappa t/2}$, and for the second
cavity we apply its time reversal, that is, an exponentially decreasing
pulse with the decay rate $\kappa /2$. The pulse duration $T$ should satisfy
$\kappa T\gg 1$, and the initial value $\widetilde{\varepsilon }\left(
0\right) $ is determined by the requirement $r_{o}\alpha \left( T/2\right) =1
$. The single-photon pulse connecting the two cavities then has a time
symmetric shape with $f\left( t\right) \propto
\mathop{\rm sech}%
\left[ \kappa \left( t-T/2\right) /2\right] $.

\subsection{Entanglement generation and atomic quantum teleportation}

If one has two cavities, each with an atom inside, one can maximally
entangle these two atoms 1 and 2 by the following method: The two atoms are
initially prepared in the state $\left| g\right\rangle $, and then we excite
them to the state $\left| s\right\rangle $ with a small possibility $%
p_{0}\approx 1-\exp \left( -\kappa \int_{0}^{T}\sin ^{2}\theta \left( \tau
\right) d\tau \right) $ through an incomplete adiabatic passage. The output
pulses from the two cavities, each with a mean photon number $p_{0}$, have a
definite pulse shape as we have shown before so that they can interfere with
each other at a $50\%-50\%$ beam splitter. The outputs of the beam splitter
are detected by two single-photon detectors, and if we register a photon
from one of the detectors, due to the interference, we do not know from
which cavity the registered photon comes from. The two atoms 1 and 2 are
thus projected to a quantum superposition state $\left( \left|
g\right\rangle _{1}\left| s\right\rangle _{2}\pm \left| s\right\rangle
_{1}\left| g\right\rangle _{2}\right) /\sqrt{2}$, which is maximally
entangled. The method described here is just an adiabatic passage version of
the scheme in Refs. \cite{9,10}. By transformation from the Raman version to
the adiabatic passage version, the output pulse shapes become insensitive to
the random atom position as is required for interference, which is important
for the scheme to work with hot atoms.

After entanglement has been generated, one can use it for atomic Bell
inequality detection, for quantum teleportation of atomic states \cite{10},
or even for realization of quantum repeaters \cite{26}. To realize quantum
repeaters, what one needs to do is to simply replace the atomic ensemble in
the scheme in Ref. \cite{26} by the setup of a single atom in a high-Q
cavity.

For the above applications, in addition to the entanglement generation, we
also need to do some single-bit operations. These single-bit operations
should also be performed in a suitable way so that they are insensitive to
the random atom position ${\bf r}$. One way is to still use adiabatic
passages. It is possible to realize any single-bit operation with adiabatic
passages \cite{27,28}, but for this purpose one needs to use a four-level
scheme instead of the $\Lambda $ configuration. There is actually a simpler
way for getting robust single-bit operations based on the Raman transitions.
Note that for single-bit operations, we do not need to use any cavity mode
or cavity effect. We can shine two travelling-wave beams on the atom
respectively coupling to the transitions $\left| g\right\rangle \rightarrow
\left| e\right\rangle $ and $\left| s\right\rangle \rightarrow \left|
e\right\rangle $. They are assumed to be collinear and propagating along the
$x$ axis which is perpendicular to the cavity axis $z$. The two
travelling-wave beams are broad with the beam radius much lager than the
typical variation length of the atom position. With this condition, the two
Rabi frequencies for the transitions $\left| g\right\rangle \rightarrow
\left| e\right\rangle $ and $\left| s\right\rangle \rightarrow \left|
e\right\rangle $ are given by $\Omega _{1}\left( {\bf r}\right) =\Omega
_{10}e^{i\omega _{ge}x/c}$ and $\Omega _{2}\left( {\bf r}\right) =\Omega
_{20}e^{i\omega _{se}x/c}$, respectively, where $\Omega _{10}$ and $\Omega
_{20}$ are basically independent of the atom position ${\bf r}$. Under a
lager detuning $\Delta $, the effective Raman coupling rate $\Omega _{R}\sim
\Omega _{1}\left( {\bf r}\right) \Omega _{2}^{\ast }\left( {\bf r}\right)
/\Delta \propto e^{i\omega _{gs}x/c}$, which is very insensitive to the
random atom position ${\bf r}$ since $c/\omega _{gs}$\ is typically much
larger the variation length of the position. Therefore, as long as we do not
need to use the cavity effect, a Raman scheme with two broad
collinearly-propagating beams suffice to eliminate the sensitivity to the
random atom position.

\subsection{Quantum computation}

In principle, we can also use this setup for quantum computation \cite{11},
and eliminate the requirement of the Lamb-Dick condition by performing all
the quantum gates using adiabatic passages \cite{27,29} with appropriate
configurations. However, the requirements for a universal quantum
computation are more challenging compared with the applications mentioned
above, and this it is somewhat a long-term goal, so we do not discuss here
the details of this possibility.

\section{Discussion of the experimental situation}

Finally, let us mention the current experimental situation related to this
work at the Caltech group. In the Caltech experiment, a single cesium atom
is trapped inside the high-finesse cavity with a FORT beam. The atomic
states $\left| g\right\rangle $, $\left| s\right\rangle $, and $\left|
e\right\rangle $ correspond to the hyperfine levels $\left(
6S_{1/2},F=3,m=+3\right) $, $\left( 6S_{1/2},F=4,m=+4\right) $, and $\left(
6P_{3/2},F=4,m=+4\right) $, respectively. The FORT beam is incident on one
of the cavity mirrors and resonant to a longitudinal mode of the cavity.
Presently, the FORT wavelength $\lambda _{\text{FORT}}$ is $936$ nm. This
wavelength was chosen because with such a beam, the trapping potentials for
the ground $6S_{1/2}$ manifold and the excited $6P_{3/2}$ manifold are
nearly identical. Considering only this reduced manifold of states, we find
that the expression for the FORT potential of the ground states $\left|
g\right\rangle $ and $\left| s\right\rangle $ is given by \cite{wieman}
\begin{equation}
U_{FORT}({\bf r})=\frac{\pi c^{2}\gamma _{s}}{2\omega _{0}^{3}}(\frac{2}{%
\Delta _{2}}+\frac{1}{\Delta _{1}})I({\bf r}).
\end{equation}%
Here, $\Delta _{1}(\Delta _{2})$ is the detuning of the FORT light of
frequency $\omega _{\text{FORT}}=2\pi c/\lambda _{\text{FORT}}$ from the $%
P_{1/2}(P_{3/2})$ level, and $\gamma _{s}/2\pi \approx 5.2$ MHz is the
spontaneous decay rate of the level $6P_{3/2}$. The intensity $I({\bf r})$
of the standing wave mode inside the cavity is given by
\begin{equation}
I({\bf r})=\frac{8P}{\pi w_{0}^{2}}\sin ^{2}(\frac{2\pi z}{\lambda _{\text{%
FORT}}})\exp (-\frac{x^{2}+y^{2}}{w_{0}^{2}})\text{,}
\end{equation}%
where $w_{0}\approx 25$ $\mu $m is the waist of the Gaussian mode, and $P$
is the power of the FORT beam inside the cavity. The trap frequencies $\nu _{%
\text{axial}},$ $\nu _{\text{radial}}$ in the axial and radial directions
follow from these expressions as
\begin{equation}
(\nu _{\text{axial}},\nu _{\text{radial}})=\frac{1}{2\pi \hbar }(\sqrt{2U_{0}%
\frac{\hbar ^{2}\omega _{FORT}^{2}}{mc^{2}}},\sqrt{2U_{0}\frac{\hbar ^{2}}{%
m(w_{0})^{2}}}),
\end{equation}%
where $U_{0}=U_{FORT}({\bf 0})$ is the trap depth. The typical power of the
FORT beam measured outside the cavity is about $1$ mW, and the power $P$
inside the cavity is enhanced by a factor of the cavity finesse, which is
about $2200$ at the wavelength of the FORT beam. With this number, the
typical values for the trap depth and frequencies are given by $U_{0}\approx
38$ MHz, $\nu _{\text{axial}}\approx 510$ kHz, and $\nu _{\text{radial}}$ $%
\approx 4.3$ kHz, respectively. The current achievable temperature $T_{\text{%
tem}}$ of the trapped atom is a significant fraction of the trap depth $U_{0}
$ (such as a half). With such a temperature, the spatial extent of the
atomic motion in the axial and radial directions are estimated respectively
by

\begin{equation}
\delta z/\lambda _{\text{FORT}}\approx \left( 1/2\pi \right) \arcsin \sqrt{%
k_{B}T_{\text{tem}}/U_{0}},
\end{equation}%
\begin{equation}
\delta r_{\perp }\approx w_{0}\sqrt{-\ln (1-k_{B}T_{\text{tem}}/U_{0})},
\end{equation}%
which will induce significant variation of the coupling rate $g\left( {\bf r}%
\right) $ given by Eq. (1). For example, for the temperature of half of the
trap depth, the axial uncertainty is $120$ nm, while the radial one is $15$ $%
\mu $m. These uncertainties cause variations in $g$ of $30\%$ due to the
radial motion, and $35\%$ due to the axial one. Therefore, within the
current experimental technique, it is important to use the method given in
this paper to make the application schemes insensitive to the variation of $%
g\left( {\bf r}\right) $. The time scale for the variation of $g\left( {\bf r%
}\right) $ is estimated by the inverse of the trap frequencies $\nu _{\text{%
axial}},\nu _{\text{radial}}$ in the axial and radial directions,
respectively. The operation time $T$ is typically significantly shorter than
$1/\nu _{\text{radial}}$, but longer or comparable to $1/\nu _{\text{axial}}$%
. So, we can take static average of $g\left( {\bf r}\right) $ in the radial
direction, and dynamical average of $g\left( {\bf r}\right) $ in the axial
direction as discussed in the note \cite{19'}.

We also would like to note that although the method in this paper shows that
many application schemes of the cavity QED setup can be demonstrated before
the achievement of efficient cooling of the trapped atom inside the cavity,
the cooling is still an important and desirable technology yet to achieve to
significantly increase the trapping time of the atom. In addition, a
combination of the cooling technology and the method here could further
improve the performance of various application schemes.

\section{Summary}

In summary, we have shown that the setup with a single trapped atom in a
high-Q cavity can be used to realize many cavity QED and quantum information
processing schemes even if the atom is still hot and not fully localized in
space (the Lamb-Dick condition is not yet satisfied). This could
significantly simplify the on-going experiments since it means many
interesting schemes can be demonstrated with the present technology before
the achievement of efficient cooling inside the cavity. Even with further
advances in atomic localization in cavity QED, our scheme should lead to a
greater robustness again certain experimental nonidealities. The basic idea
of this method is to design an appropriate adiabatic passage so that the
relevant dynamics only depend on the ratio of two coupling rates. Though
each of the coupling rates is sensitive to the unknown or time-varying atom
position, their ratio is fixed and controllable as the two rates depend on
the random atom position in the same way with the appropriate interaction
configuration that we have described. We confirm the validity of this method
by solving the complete model which describes the realistic setup. The
approximate analytical solution and the exact numerical simulations agree
with each other. From the numerical simulations, we also calculate
quantitatively various noise magnitudes in this setup, and show one can
achieve reasonably good performance with the values of the parameters based
on the present technology. Finally, we show that this method can be
incorporated into many previous schemes, allowing the demonstration of these
application schemes without the requirement of the full localization of the
atom.

{\bf Acknowledgments}: This work was supported by the Caltech MURI Center
for Quantum Networks under ARO Grant No. DAAD19-00-1-0374, by the National
Science Foundation under Grant No. EIA-0086038, and by the Office of Naval
Research. L.M.D. also acknowledge support from the Chinese Science
Foundation, Chinese Academy of Sciences, and the national "97.3" project.


\begin{references}
\bibitem{19ab} A leakage error means that after a spontaneous emission, the
system goes out of the Hilbert space spanned by the state (10) with all the
possible coefficients, since in this case one has a photon in a spontaneous
emission mode.
\end{references}
\end{document}